\documentclass[11pt]{article}
\usepackage[dvips]{graphicx}

\begin{document}
\title{Overall Feature of CP dependence
for Neutrino Oscillation Probability
in Arbitrary Matter Profile}
\author{
\sc 
{Hidekazu Yokomakura$^1$}
\thanks{E-mail address:yoko@eken.phys.nagoya-u.ac.jp} 
{, Keiichi Kimura$^1$}\thanks
{E-mail address:kimukei@eken.phys.nagoya-u.ac.jp} 
{, and Akira Takamura$^{1,2}$}
\thanks{E-mail address:takamura@eken.phys.nagoya-u.ac.jp} 
\\
\\
\\
{\small \it $^1$Department of Physics, Nagoya University,}
{\small \it Nagoya, 464-8602, Japan}\\
{\small \it $^2$Department of Mathematics, 
Toyota National Collage of Technology}\\
{\small \it Eisei-cho 2-1, Toyota-shi, 471-8525, Japan}}
\date{}
\maketitle
\vspace{-10cm}
\begin{flushright}
  {DPNU 02-21}\\
  {July 2002}
\end{flushright}
\vspace{10.5cm}
\vspace{-2.5cm}
\vspace{1cm}
\begin{abstract}
We study the CP dependence of neutrino oscillation probability
for all channels in arbitrary matter profile 
within three generations. 
We show that an oscillation probability 
for $\nu_e \to \nu_\mu$ can be written in the form 
$P(\nu_e \to \nu_\mu)
=A_{e\mu} \cos \delta + B_{e\mu} \sin \delta + C_{e\mu}$
without any approximation
using the CP phase $\delta$. 
This result holds not only in constant matter 
but also in arbitrary matter. 
Another probability for $\nu_\mu \to \nu_\tau$
can be written in the form  
$P(\nu_\mu \to \nu_\tau)=
   A_{\mu\tau} \cos \delta
 + B_{\mu\tau} \sin \delta
 + C_{\mu\tau}
 + D_{\mu\tau} \cos 2\delta
 + E_{\mu\tau} \sin 2\delta
$.  
The term which is proportional to $\sin 2\delta$ 
disappear, namely $E_{\mu\tau}=0$, in symmetric matter. 
It means that the probability reduces to  
the same form as in constant matter. 
As for other channels, probabilities in arbitrary matter are at most 
the quadratic polynomials of 
$\sin \delta$ and $\cos \delta$ as in the above two channels. 
In symmetric matter, the oscillation probability
for each channel reduces to the same form with respect to $\delta$ 
as that in constant matter. 
\end{abstract}

\section{Introduction}
\hspace*{\parindent}
In solar and atmospheric neutrino experiments, 
the $\nu_e$ deficit
\cite{solar}
and the $\nu_\mu$ anomaly
\cite{atm} have been observed. 
These results strongly suggest the finite mixing angles 
$\theta_{12}$ and $\theta_{23}$ 
and the finite mass squared differences
$\Delta_{12}$ and $\Delta_{23}$, 
where 
$\Delta_{ij}=m^2_i-m^2_j$.
Within the framework of three generations, 
there are two more parameters $\theta_{13}$ and $\delta$ 
to be determined. 
About $\theta_{13}$, only upper bound is obtained 
from CHOOZ experiment \cite{CHOOZ} and the information on 
the CP phase $\delta$ 
is not obtained at all. 
In order to determine these parameters, 
several long baseline experiments using artificial neutrino beam
will be planned
\cite{lbl}, 
and it is important to study the effect when 
the neutrino pass through
the matter
\cite{MSW}. 
The main physics goal in these experiments 
is to measure the value of $\delta$. 
In this paper, we study the CP dependence 
of oscillation probability for all channels 
in arbitrary matter profile. 

Before giving our results, 
let us review the works 
on CP violation in three neutrino oscillation. 
At first we introduce the CP-odd asymmetry
$
\Delta P_{\alpha\beta}^{CP}
= P(\nu_\alpha \to \nu_\beta)
-P(\bar{\nu}_\alpha \to \bar{\nu}_\beta)
$.
In disappearance channel, 
$
\Delta P_{\alpha\alpha}^{CP}
$
is exactly equal to $0$ in vacuum and 
independent of $\delta$. 
However, 
$
\Delta P_{\alpha\alpha}^{CP}
$
is not always equal to $0$
in matter. 
This is due to the genuine CP violation and/or 
fake CP violation from matter effects. 
In the case of $\alpha=e$, 
Kuo and Pantaleone \cite{Kuo}
have shown that 
$P(\nu_e \to \nu_e)$
does not depend on $\delta$ in the context of solar neutrino problem. 
Therefore, 
$
\Delta P_{ee}^{CP}\neq 0
$ 
arises from matter effects
\footnote{
Minakata and Watanabe have shown that 
$P(\nu_e \to \nu_e)$
slightly depends on $\delta$  
if we take into account the loop correction
even in the standard model.
In this paper, we do not consider the loop correction 
as these effects are safely neglected \cite{Minakata-Watanabe}.}. 
However, in the case of $\alpha=\mu$, $P(\nu_\mu \to \nu_\mu)$ 
has the CP-odd term in asymmetric matter profile as pointed 
out by Minakata and Watanabe 
\cite{Minakata-Watanabe}.
We investigate the CP dependence in more detail 
in this paper.
 
Let us consider the appearance channels.
As the CP-odd asymmetry is proportional to $\sin\delta$ 
in vacuum, 
$
\Delta P_{\alpha\beta}^{CP} \neq 0
$
means that the discovery of CP violation.
However, the situation completely changes 
when the matter effects are taken into account. 
Namely, 
$
\Delta P_{\alpha\beta}^{CP}\neq 0
$
does not always mean the existence 
of CP violation \cite{Langacker}, 
because the fake CP violation due to matter effects exists. 
Here, 
it is difficult to separate  
genuine CP violation
due to $\delta$
from 
fake CP violation. 
One of the methods to solve these problems is to take into account 
mass hierarchy approximation $|\Delta_{21}| \ll |\Delta_{32}|$. 
Actually, some approximate formulae are given by Arafune {\it et al.} 
at low energy region \cite{AS}
and by Cervera {\it et al.} \cite{Cervera} and Freund \cite{Freund} 
at high energy region. 

Next, we introduce the T-odd asymmetry
$
\Delta P^T_{\alpha\beta}
=P(\nu_\alpha \to \nu_\beta)-P(\nu_\beta \to \nu_\alpha)
$.
Krastev and Petcov
\cite{Krastev}
have shown that 
$\Delta P^T_{\alpha\beta}$
is proportional to 
$\sin \delta$ exactly 
in constant matter.
Recently, 
Naumov
\cite{Naumov} 
, Harrison and Scott
\cite{Harrison-Scott}
have derived
the simple identity on the Jarlskog factor 
$J$ \cite{Jarlskog}
 as 
$
\tilde{\Delta}_{12}
\tilde{\Delta}_{23}
\tilde{\Delta}_{31}
\tilde{J}
=
{\Delta}_{12}
{\Delta}_{23}
{\Delta}_{31}
{J}
$, where quantities with tilde 
represent those in matter. 
We can simply understand that $\Delta P^T_{\alpha\beta}$ 
is proportional to $\sin\delta$ from this identity.
We have studied  
the matter enhancement of
$\tilde{J}$
\cite{YKT} taking advantage of this identity. 
Furthermore, 
Parke and Weiler
have investigated 
the matter enhancement of the $\Delta P^T_{e\mu}$ 
\cite{Parke-Weiler}. 

There are some works on the deviation from constant matter.
In long baseline experiments, 
we need to estimate the validity of constant density
approximation because the earth matter density largely 
changes along to the path of neutrino. 
The matter profile of the earth
is approximately expressed
by Preliminary Reference Earth Model(PREM)
\cite{PREM}.
Minakata and Nunokawa\cite{Minakata9705}
give the oscillation probability using mass hierarchy 
and adiabatic approximations.
For the distance less than $L=3000$ km, 
the matter density fluctuation is small 
and the constant density approximation is valid. 
On the other hand, it has been shown that
the fluctuation of the density
cannot be ignored for the distance greater than $L=7000$km
\cite{KS98K2K,OS,Miura0106,Pinney}. 
Furthermore, the constant density 
approximation is not valid 
in the case that  
the matter density profile
is different from PREM
and the asymmetric part exists. 
It is pointed out that $\Delta P^{T}_{\alpha\beta}$ has the term proportional 
to $\cos \delta$ in arbitrary matter. 
\cite{Miura0102,Akhmedov}. 
We investigate this feature in more detail in this paper.

In previous paper, 
we have proposed the new method applicable to constant matter.   
This method is to estimate the product of effective 
Maki-Nakagawa-Sakata matrix 
elements\cite{MNS} 
$\tilde{U}_{\alpha i}\tilde{U}_{\beta i}^{*}$
without directly calculating
$\tilde{U}_{\alpha i}$. 
We have shown that the oscillation probability $P(\nu_e \to \nu_\mu)$
is written in the linear combination
\footnote{
It is not so easy to obtain our result
from the effective mixing and effective CP phase
given by Zaglauer and Schwarzer \cite{Zaglauer}} 
of $\cos \delta $
and
$\sin \delta $
 exactly
\cite{Kimura0203} 
\begin{eqnarray}
\label{eq:kty1}
P(\nu_e \to \nu_\mu)=
   A_{e\mu} \cos \delta
 + B_{e\mu} \sin \delta
 + C_{e\mu}, 
\end{eqnarray}
in constant matter.
In other channels, 
for example, 
\begin{eqnarray}
\label{eq:kty2}
P(\nu_\mu \to \nu_\tau)=
   A_{\mu\tau} \cos \delta
 + B_{\mu\tau} \sin \delta
 + C_{\mu\tau}
 + D_{\mu\tau} \cos 2\delta
. 
\end{eqnarray}
It is found that the probability is quadratic polynomial
of 
$\cos \delta $, 
$\sin \delta $, 
and the CP dependence was equal to in vacuum
\cite{Kimura0205}. 

In this paper, 
we give the exact CP dependence of oscillation probability 
for all channels in arbitrary matter profile. 
For the purpose, we decompose the Hamiltonian $H$
in the form $H=(O_{23}\Gamma_\delta)H'(O_{23}\Gamma_\delta)^\dagger$
using 2-3 rotation matrix $O_{23}$ and CP phase matrix $\Gamma_\delta
={\rm diag}(1,1,e^{i\delta})$.
This decomposition plays a key role in our paper.
As a result, we obtain the probability  
for $\nu_e \to \nu_\mu$ as 
\begin{equation}
P(\nu_e \to \nu_\mu)
=  A_{e\mu} \cos \delta
 + B_{e\mu} \sin \delta
 + C_{e\mu}.
\end{equation}
This has the same form with respect to $\delta$ 
as in eq.(\ref{eq:kty1}) in constant matter. 
On the other hand, 
for $\nu_\mu \to \nu_\tau$, 
we show that the probability is given by
\begin{equation}
\label{eq:mutau}
P(\nu_\mu \to \nu_\tau)
=
   A_{\mu\tau} \cos \delta
 + B_{\mu\tau} \sin \delta
 + C_{\mu\tau}
 + D_{\mu\tau} \cos 2\delta
 + E_{\mu\tau} \sin 2\delta.
\end{equation}
Comparing the equations (\ref{eq:mutau}) with (\ref{eq:kty2}), 
the probability in arbitrary matter profile
has the term proportional to $\sin 2 \delta$ 
which does not exist in the probability for constant matter. 
Furthermore, in the case of symmetric matter profile, 
we show that this additional term disappears, namely $E_{\mu\tau}=0$, 
and the probability reduces to the same form as 
in constant matter. 

\section{CP Dependence in Arbitrary Matter Profile}
\hspace*{\parindent}
In this section, we study the exact CP dependence of neutrino 
oscillation probability
in arbitrary matter profile. 
The Schr\"{o}dinger equation 
for neutrino is 
\begin{eqnarray}
  \label{eq:Schrodinger}
  i \frac{\partial \nu }{\partial t} = H \nu, 
\end{eqnarray}
where $H$ is the Hamiltonian in matter and $\nu$ 
is flavor eigenstate $\nu=(\nu_e, \nu_\mu, \nu_\tau)^T$. 
We introduce the MNS matrix 
which relates the flavor eigenstate $\nu_\alpha$
to the mass eigenstate $\nu_i$.
The MNS matrix $U$ in the standard parametrization is represented as
\begin{eqnarray}
  \label{eq:mns-standard}
  U_{}=O_{23}\Gamma_{\delta}O_{13}\Gamma_{\delta}^{\dagger}O_{12}, 
\end{eqnarray}
where
 $\Gamma_{\delta}=\mbox{diag}(1,1,e^{i\delta})$ 
and 
\begin{eqnarray}
O_{23} = \left(
    \begin{array}{ccc}
 1 & 0 & 0 \\
 0 & c_{23} & s_{23} \\
 0 & - s_{23} & c_{23}
    \end{array}
  \right),
\end{eqnarray}
using the abbreviation $s_{ij}=\sin\theta_{ij}$ and 
 $c_{ij}=\cos\theta_{ij}$.
$O_{13}$ and $O_{12}$ represent 1-3 and 1-2 rotation matrix 
like $O_{23}$, respectively.
By using this relation (\ref{eq:mns-standard}), 
we can rewrite the $H$ as
 \begin{eqnarray}
  \label{eq:Freund}
 H
&=&
 \frac{1}{2E}
 \left[
U 
{\rm diag}(0, \Delta_{21}, \Delta_{31})
U^{\dagger}
+  {\rm diag}(a(t), 0, 0)
 \right] \\
&=&
 \frac{1}{2E}
 O_{23}\Gamma_{\delta}
 \left[
 O_{13}O_{12} {\rm diag}(0, \Delta_{21}, \Delta_{31})
 O_{12}^T O_{13}^T
 + {\rm diag}(a(t), 0, 0)
 \right]
 \Gamma_{\delta}^\dagger O_{23}^T,
  \label{2}
 \end{eqnarray}
where $a(t)$ is matter potential defined by 
$a(t)=2 \sqrt2 G_F N(t)_e E$, and $G_F$, $N(t)_e$, $E$ are 
respectively Fermi constant, electron number density and 
neutrino energy.
This equation (\ref{2}) means that 
the Hamiltonian can be decomposed into two parts.
One is 1-2 and 1-3 mixing part 
which contain matter effects.
The other is 2-3 mixing and CP phase $\delta$ part which does not 
contain matter effects.
It is noted that this decomposition is garanteed by the relation
\begin{eqnarray}
O_{23}\Gamma_{\delta}
{\rm diag}(a(t), 0, 0)
(O_{23}\Gamma_{\delta})^\dagger
={\rm diag}(a(t), 0, 0). 
 \label{23cpinv}
\end{eqnarray}
We can separate CP phase $\delta$ from matter effects
by taking advantage of this decomposition (\ref{2}). 
Changing $\nu$ to $\nu'$ as  
\begin{eqnarray}
  \label{eq:base}
  \nu' = 
(O_{23}\Gamma_{\delta})^\dagger
\nu
,
\end{eqnarray}
the Schr\"{o}dinger equation
(\ref{eq:Schrodinger})
is rewritten as 
\begin{eqnarray}
  \label{eq:b-schrodinger}
  i \frac{\partial \nu' }{\partial t} = H' \nu',
\end{eqnarray}
where 
\begin{eqnarray}
{H'}&=&
  \frac{1}{2E}
\left[
O_{13}O_{12}
{\rm diag}(0, \Delta_{12}, \Delta_{13})
 (O_{13}O_{12})^T 
+
{\rm diag}(a(t), 0, 0)
\right].
 \label{3}
\end{eqnarray}
We emphasise that the reduced Hamiltonian $H'$
does not contain the 2-3 mixing and CP phase
and is real symmetric.
For anti-neutrino, we obtain the similar results by 
the replacements $\delta \to -\delta$ 
and in $a(t) \to -a(t)$. 

Next, we introduce the time evolution operator $S(t)$ and $S'(t)$
which is defined by the solution of the Schr\"{o}dinger equation 
\begin{eqnarray}
   \label{eq:b-schrodinger-s}
   \nu(t) = S(t)\nu(0), \quad
   \nu'(t) = S'(t)\nu'(0).
\end{eqnarray}
The relation between $S(t)$ and $S'(t)$ is determined by 
the transformation (\ref{eq:base}) and is given by
\begin{eqnarray}
  \label{eq:transform-s}
  S(t)
  =
(O_{23}\Gamma_{\delta})
  S'(t)
(O_{23}\Gamma_{\delta})^\dagger.
\end{eqnarray}
By taking the component of (\ref{eq:transform-s}), 
the relation between the time evolution operators 
for each flavour is given by  
\begin{eqnarray}
  \label{eq:general-result-amp-ee}
&&
  S_{ee}=  S'_{ee}, 
\\
&&
  S_{\mu e}=  S'_{\mu e} c_{23} + S'_{\tau e} s_{23} e^{i\delta}, 
\\
&&
  S_{\tau e}=  -S'_{\mu e} s_{23} + S'_{\tau e} c_{23} e^{i\delta}, 
\\
&&
  S_{\mu \mu} =   S'_{\mu \mu}   c_{23}^2
                + S'_{\mu \tau}  c_{23} s_{23} e^{-i\delta}
                + S'_{\tau \mu}  c_{23} s_{23} e^{i\delta} 
                + S'_{\tau \tau} s_{23}^2 , 
\\
&&
  S_{\tau \mu}  = - S'_{\mu \mu}   c_{23} s_{23}
                  - S'_{\mu \tau}  s_{23}^2 e^{-i\delta}
                  + S'_{\tau \mu}  c_{23}^2 e^{i\delta} 
                  + S'_{\tau \tau} c_{23} s_{23} , 
\\
&&
  \label{eq:general-result-amp-tautau}
  S_{\tau \tau} =   S'_{\mu \mu}   s_{23}^2
                  - S'_{\mu \tau}  c_{23} s_{23} e^{-i\delta}
                  - S'_{\tau \mu}  c_{23} s_{23} e^{i\delta} 
                  + S'_{\tau \tau} c_{23}^2, 
\end{eqnarray}
Here, $S_{\alpha\beta}$ represents 
the transition amplitude for $\nu_\beta \to \nu_\alpha$. 
$S_{e \mu}$, $S_{e \tau}$
and 
$S_{\mu \tau}$ are 
obtained from 
$S_{\mu e}$, 
$S_{\tau e}$ 
and  
$S_{\tau \mu}$ 
respectively 
by the replacements  
$S'_{\alpha\beta} \to S'_{\beta\alpha}$, $\delta \to -\delta$. 
Substituting 
(\ref{eq:general-result-amp-ee})-(\ref{eq:general-result-amp-tautau})
 into the relation 
\begin{eqnarray}
P(\nu_\alpha \to \nu_\beta)=|S_{\beta\alpha}|^2  \label{PS}, 
\end{eqnarray}
the oscillation probabilities in arbitrary matter profile 
are given by
\begin{eqnarray}
  \label{eq:general-result-prob-ee}
&&
  P(\nu_e \to \nu_e) =  C_{ee}, 
\\
&&
  \label{eq:general-result-prob-emu}
  P(\nu_e \to \nu_\mu) =
     A_{e\mu} \cos\delta
  +  B_{e\mu} \sin\delta
  +  C_{e\mu}, 
\\
&&
  \label{eq:general-result-prob-etau}
  P(\nu_e \to \nu_\tau) =
     A_{e\tau} \cos\delta
  +  B_{e\tau} \sin\delta
  +  C_{e\tau}, 
\\
&&
  P(\nu_\mu \to \nu_\mu) =
     A_{\mu\mu} \cos\delta
  +  B_{\mu\mu} \sin\delta
  +  C_{\mu\mu}
  +  D_{\mu\mu} \cos2\delta
  +  E_{\mu\mu} \sin2\delta
, 
  \label{eq:general-result-prob-mu-mu}
\\
&&
  P(\nu_\tau \to \nu_\tau) =
     A_{\tau\tau} \cos\delta
  +  B_{\tau\tau} \sin\delta
  +  C_{\tau\tau}
  +  D_{\tau\tau} \cos2\delta
  +  E_{\tau\tau} \sin2\delta
, 
  \label{eq:general-result-prob-tau-tau}
\\
&&
  P(\nu_\mu \to \nu_\tau) =
     A_{\mu\tau} \cos\delta
  +  B_{\mu\tau} \sin\delta
  +  C_{\mu\tau}
  +  D_{\mu\tau} \cos2\delta
  +  E_{\mu\tau} \sin2\delta
, 
  \label{eq:general-result-prob-mu-tau}
\end{eqnarray}
and the other probabilities
$P(\nu_\mu \to \nu_e)$, 
$P(\nu_\tau \to \nu_e)$ and 
$P(\nu_\tau \to \nu_\mu)$
are obtained by the replacements 
$S'_{\alpha\beta} \to S'_{\beta\alpha}$ and $\delta \to -\delta$
in $P(\nu_e \to \nu_\mu)$, 
$P(\nu_e \to \nu_\tau)$ and 
$P(\nu_\mu \to \nu_\tau)$, respectively. 
Here all coefficients
$A_{e\mu},\cdots, E_{\mu \tau}$
are constructed from the mixing angle
$\theta_{23}$ and 
$S'$ including matter effects. 
See appendix \ref{sec:appendix-a} for detail.
The oscillation probabilities  for ''anti-neutrino'' 
are also obtained by the replacements $\delta \to -\delta$
and $a(t) \to -a(t)$. 

For these 
eqs.(\ref{eq:general-result-prob-ee})-(\ref{eq:general-result-prob-mu-tau}), 
we emphasize the following three points. 
First, the survival probability $P(\nu_e \to \nu_e)$ 
in eq.(\ref{eq:general-result-prob-ee})
is compretely independent of CP phase $\delta$.
This coincides with the result 
by Kuo and Pantaleone \cite{Kuo}. 
Second, the transition probabilities $P(\nu_e \to \nu_{\mu})$ 
and $P(\nu_e \to \nu_{\tau})$ in 
eq.(\ref{eq:general-result-prob-emu})
and 
eq.(\ref{eq:general-result-prob-etau})
are linear polynomials of $\sin \delta$ and $\cos \delta$. 
These features coincide with the results in
constant matter 
\cite{Kimura0203,Kimura0205}.
Third, $P(\nu_\mu \to \nu_\tau)$, 
$P(\nu_\mu \to \nu_\mu)$
and
$P(\nu_\tau \to \nu_\tau)$ in 
eq.(\ref{eq:general-result-prob-mu-mu})-(\ref{eq:general-result-prob-mu-tau})
are at most quadratic polynomials of 
$\sin \delta$ and $\cos \delta$. 

We also comment the CP trajectory introduced by Minakata and Nunokawa.
This is an orbit in the bi-probability space
when 
$\delta$ 
changes from $0$ to $2\pi$
\cite{Minakata0204,Minakata0108}. 
Eq.(\ref{eq:general-result-prob-ee}) shows that
CP trajectory is exactly elliptic even in arbitrary 
matter profile. 
In addition, we point out that 
the dependence of $\theta_{23}$
for the oscillation probabilities
is completely understood from 
eqs.(\ref{eq:general-result-prob-ee})-(\ref{eq:general-result-prob-tau-tau}).
See appendix \ref{sec:appendix-a} for detail.

It is also noted that there are two features in  
asymmetric matter profile. 
First, the terms proportional to
$\sin \delta$ and $\sin 2\delta$ are appeared 
in $P(\nu_\mu \to \nu_\mu)$ and $P(\nu_\tau \to \nu_\tau)$. 
The term proportional to $\sin 2\delta$ 
are appeared in $P(\nu_\mu \to \nu_\tau)$ and 
$P(\nu_\tau \to \nu_\mu)$. 
These terms do not exist in constant matter
\cite{Kimura0203,Kimura0205}. 
Second, $\Delta P^{T}_{\alpha\beta}$
is not proportional to $\sin \delta$ in asymmetric matter 
as in constant matter \cite{Krastev}. 
In the next section, we describe these features in more detail.

\section{CP Dependence in Symmetric Matter Profile}
\label{sec:sym-prof}
\hspace*{\parindent}
In this section,  
we study the CP dependence of $P(\nu_\alpha \to \nu_\beta)$ 
in symmetric matter profile 
as special case of the previous section. 
In the case of symmetric matter 
along neutrino path, 
the time evolution operator $S'$ becomes symmetric matrix
\begin{eqnarray}
  \label{eq:direct-reverse-S-dash-sym}
  S_{\alpha\beta}'
  =
  S_{\beta\alpha}', 
\end{eqnarray}
for flavour indices
\cite{Akhmedov,Fishbane}. 
As results, 
the relations between the coefficients of 
$P(\nu_{\alpha} \to \nu_e)$ 
 eq.(\ref{eq:general-result-prob-ee})-(\ref{eq:general-result-prob-etau})
 and $P(\nu_e \to \nu_{\alpha})$ 
are given by 
\begin{eqnarray}
  \label{eq:abc}
&&
  A_{\mu e}=A_{e\mu}
, 
\hspace{1em}
  B_{\mu e}=-B_{e \mu}
, 
\hspace{1em}
  C_{\mu e}=C_{e \mu},
\\ &&
  A_{\tau e}=A_{e \tau}
, 
\hspace{1em}
  B_{\tau e}=-B_{e \tau}
, 
\hspace{1em}
  C_{\tau e}=C_{e \tau}. 
\end{eqnarray}
See appendix \ref{sec:appendix-b} for detail calculation.  
The probabirity $P(\nu_e \to \nu_{\alpha})$
have the same form with respect to $\delta$ 
as 
eqs.(\ref{eq:general-result-prob-ee})-(\ref{eq:general-result-prob-etau})
in arbitrary matter profile.

On the other hand, 
applying the condition (\ref{eq:direct-reverse-S-dash-sym})
to the probability
 (\ref{eq:general-result-prob-mu-mu})-(\ref{eq:general-result-prob-mu-tau}), 
we obtain the remarkable relations
\begin{equation}
    \label{eq:be}
B_{\mu \mu}=B_{\tau \tau}
=E_{\mu \mu}=E_{\tau \tau}
=E_{\mu \tau}=E_{\tau \mu}
=0, 
\end{equation}
where the detailed calculation is given 
in the appendix \ref{sec:appendix-b}.
Using these relations, 
the oscillation probabilities
 (\ref{eq:general-result-prob-mu-mu})-(\ref{eq:general-result-prob-mu-tau})
have more simple form such as
\begin{eqnarray}
&&
  \label{eq:symmetric-result-prob--sym-mumu}
  P(\nu_\mu \to \nu_\mu) =
     A_{\mu\mu} \cos\delta
  +  C_{\mu\mu}
  +  D_{\mu\mu} \cos2\delta
, 
\\
&&
  \label{eq:symmetric-result-prob--sym-tautau}
  P(\nu_\tau \to \nu_\tau) =
     A_{\tau\tau} \cos\delta
  +  C_{\tau\tau}
  +  D_{\tau\tau} \cos2\delta
, 
\\
  \label{eq:symmetric-result-prob--sym-mutau}
&&
  P(\nu_\mu \to \nu_\tau) =
     A_{\mu\tau} \cos\delta
  +  B_{\mu\tau} \sin\delta
  +  C_{\mu\tau}
  +  D_{\mu\tau} \cos2\delta
, 
\end{eqnarray}
and $P(\nu_\tau \to \nu_\mu)$
is 
simply obtained by 
replacements
$A_{\alpha\beta}$,
$B_{\alpha\beta}$,
$C_{\alpha\beta}$
and 
$D_{\alpha\beta}$. 
Then, 
the coefficients of $P(\nu_\tau \to \nu_\mu)$
are given by 
\begin{equation}
  \label{eq:abcd}
  A_{\mu \tau}=A_{\tau \mu}, 
\hspace{1em}
  B_{\mu \tau}=-B_{\tau \mu}, 
\hspace{1em}
  C_{\mu \tau}=C_{\tau \mu}, 
\hspace{1em}
  D_{\mu \tau}=D_{\tau \mu},
\end{equation}
where 
we use the condition (\ref{eq:direct-reverse-S-dash-sym})
for eqs.(\ref{eq:abcd})
 or the unitarity for last equation. 
Here, the point is that 
the term proportinal to $\sin 2\delta$ is dropped 
in eq.(\ref{eq:symmetric-result-prob--sym-mutau})
comparing with eq.(\ref{eq:general-result-prob-mu-tau}). 
The other point is that the terms proportional to $\sin \delta$
and $\sin 2 \delta$
do not exist in $P(\nu_\mu \to \nu_\mu)$
and 
$P(\nu_\tau \to \nu_\tau)$, 

As the results, the CP dependence of 
the oscillation probability
for each channel in symmetric matter
reduces to the same form as in constant matter
\cite{Kimura0205}.
This is the generalization of the result 
in our previous paper.

Finally, we study the T-odd asymmetry
$\Delta P^{T}_{\alpha\beta}
=
P(\nu_\alpha \to \nu_\beta)
-P(\nu_\beta \to \nu_\alpha)
$. 
From the unitarity relation, we easily obtain
\begin{eqnarray}
  \label{eq:deltapt}
\Delta P^{T}_{e \mu}
 =
\Delta P^{T}_{\mu \tau}
 =
\Delta P^{T}_{\tau e},  
\end{eqnarray}
and in symmetric matter profile 
we obtain 
\begin{eqnarray}
  \label{eq:deltaptemu}
\Delta P^{T}_{e \mu}
 &=&
 (
 A_{e \mu}
 -A_{\mu e}
 )
  \cos\delta
 +
 (
 B_{e \mu}
 -B_{\mu e}
 )
  \sin\delta
 +
 (
 C_{e \mu}
 -C_{\mu e}
 )
\\
  \label{eq:deltaptemu-sym}
 &=&
 2B_{e\mu}
  \sin\delta
\\
 &=&
  -4  c_{23} s_{23}
      {\rm Im}[{S}_{\mu e}^{'*} S'_{\tau e}
  ]
  \sin\delta , 
\end{eqnarray}
where we use the relations (\ref{eq:abc}) in eq.(\ref{eq:deltaptemu-sym}). 
In constant matter, Krastev and Petcov \cite{Krastev} 
have pointed out that $\Delta P^{T}_{\alpha\beta}$
is proportional to $\sin \delta$.
Our result is applicable to the symmetric matter profile, 
which corresponds to the generalization of their result, 
even if the oscillation is non-adiabatic. 

Let us turn the case of arbitrary matter profile.
$\Delta P^{T}_{\alpha\beta}$ 
in asymmetric matter is not 
proportional to $\sin\delta$ because 
the time evolution operator $S'$ 
is not symmetric, namely 
$S'_{\alpha\beta} \neq S'_{\beta\alpha}$.
More concretely speaking, the coefficients
are not symmetric for flavor indices
$A_{\alpha\beta} \neq A_{\beta\alpha}$, 
$B_{\alpha\beta} \neq - B_{\beta\alpha}$, 
$C_{\alpha\beta} \neq C_{\beta\alpha}$. 
We clarify the exact CP dependence of $\Delta P^{T}_{\alpha\beta}$ 
in asymmetric matter 
although this fact is suggested using approximation
\cite{Miura0102,Akhmedov}.

\section{Summary}
 \label{sec:summary}
\hspace*{\parindent}
We summarize the results obtained in this paper.
We have studied the CP dependence of the oscillation probability 
$P(\nu_\alpha \to \nu_\beta)$
both in arbitrary and in symmetric matter profile. 
\begin{enumerate}
\item[(i)] 
In arbitrary matter profile, 
we have found that $P(\nu_\alpha \to \nu_\beta)$ is at most 
quadratic polinomial of $\sin \delta$ and $\cos \delta$. 
The CP dependences of the probabilities can be written as 
\begin{eqnarray}
&& 
  P(\nu_e \to \nu_e) =  C_{ee}, 
\\
&&
  P(\nu_\alpha \to \nu_\beta) =
     A_{\alpha\beta} \cos\delta
  +  B_{\alpha\beta} \sin\delta
  +  C_{\alpha\beta}, 
\nonumber
\\
&&
\hspace{8em}
\mbox{ for
 ($\alpha\beta$) =
 ($e\mu$),  
 ($e\tau$),  
 ($\mu e$),  
 ($\tau e$),  
 }
\\
&&
\nonumber
  P(\nu_\alpha \to \nu_\beta) =
     A_{\alpha\beta} \cos\delta
  +  B_{\alpha\beta} \sin\delta
  +  C_{\alpha\beta}
  +  D_{\alpha\beta} \cos2\delta
  +  E_{\alpha\beta} \sin2\delta
, 
\\
&&
\hspace{8em}
 \mbox{
                  
 for
 ($\alpha\beta$) =
 ($\mu \mu$),  
 ($\mu \tau$),  
 ($\tau\mu$),  
 ($\tau\tau$).  
 }
\end{eqnarray}
\item[(ii)]
In symmetric matter profile, 
we have shown that the oscillation 
probabilities $P(\nu_e \to \nu_x)$ have the same form 
as in arbitrary matter such as  
\begin{eqnarray}
&&
  P(\nu_e \to \nu_e) =  C_{ee}, 
\\
&&
  P(\nu_e \to \nu_\mu) =
     A_{e\mu} \cos\delta
  +  B_{e\mu} \sin\delta
  +  C_{e\mu}, 
\\
&&
  P(\nu_e \to \nu_\tau) =
    A_{e \tau} \cos\delta
  +  B_{e \tau} \sin\delta
  +  C_{e\tau}.
\end{eqnarray}
Furthermore, we have shown that 
the CP dependences of other probabilities are written 
in the form as 
\begin{eqnarray}
&&
  P(\nu_\mu \to \nu_\tau) =
     A_{\mu\tau} \cos\delta
  +  B_{\mu\tau} \sin\delta
  +  C_{\mu\tau}
  +  D_{\mu\tau} \cos2\delta
, 
\\
&&
  P(\nu_\mu \to \nu_\mu) =
     A_{\mu\mu} \cos\delta
  +  C_{\mu\mu}
  +  D_{\mu\mu} \cos2\delta
, 
\\
&&
  P(\nu_\tau \to \nu_\tau) =
     A_{\tau\tau} \cos\delta
  +  C_{\tau\tau}
  +  D_{\tau\tau} \cos2\delta
.
\end{eqnarray}
It is remarkable that the 
oscillation probability for each channel 
in symmetric matter reduces to the same form 
as in constant matter. 
\end{enumerate}

\section{Acknowledgements}
\label{sec:acknowledgements}
\hspace*{\parindent}
We would like to thank Prof. A. I. Sanda 
for valuable advice. 
KK wish to thank Prof. H. Minakata and Prof. O. Yasuda 
for helpful suggestions.

\appendix
\section{Coefficients in Arbitraly Matter Profile}
\label{sec:appendix-a}
\hspace*{\parindent}
In this appendix, we give exact CP and 2-3 mixing dependences 
of the oscillation probabilities in arbitrary matter profile.
The probability for each channel is given by  
 \begin{eqnarray}
  \label{eq:ee-aa}
 \hspace{-2em}
  P(\nu_e \to \nu_e)
  &=&
  \label{eq:general-prob-ee-aa}
C_{ee} 
=
  |S'_{ee}|^2,  
\\
 \hspace{-1em}
  \label{eq:general-result-prob-emu-aa}
  P(\nu_e \to \nu_\mu) &=&
     A_{e\mu} \cos\delta
  +  B_{e\mu} \sin\delta
  +  C_{e\mu}, 
\\
  \label{eq:general-prob-emu-a-aa}
A_{e\mu} &=&   
 2 
  {\rm Re}[{S}_{\mu e}^{'*} S'_{\tau e}] c_{23} s_{23} , \hspace{0.5em}
\\
  \label{eq:general-prob-emu-b-aa}
B_{e\mu} &=& 
 - 2  
  {\rm Im}[{S}_{\mu e}^{'*} S'_{\tau e}] c_{23} s_{23} , \hspace{0.5em}
\\
  \label{eq:general-prob-emu-c-aa}
C_{e\mu} &=&  |S'_{\mu e}|^2 c_{23}^2 + |S'_{\tau e}|^2 s_{23}^2, \hspace{0.5em} 
\\
  \label{eq:general-result-prob-etau-aa}
 \hspace{-1em}
  P(\nu_e \to \nu_\tau) &=&
     A_{e\tau} \cos\delta
  +  B_{e\tau} \sin\delta
  +  C_{e\tau}, 
\\
  \label{eq:general-prob-etau-a-aa}
A_{e\tau} &=&
  - 2  
     {\rm Re}[{S}_{\mu e}^{'*} S'_{\tau e}] c_{23} s_{23},  \hspace{0.5em}
\\
  \label{eq:general-prob-etau-b-aa}
B_{e\tau} &=&
   2 
    {\rm Im}[{S}_{\mu e}^{'*} S'_{\tau e}] c_{23} s_{23}, \hspace{0.5em} 
\\
  \label{eq:general-prob-etau-c-aa}
C_{e\tau} &=&  |S'_{\mu e}|^2 s_{23}^2 + |S'_{\tau e}|^2 c_{23}^2, \hspace{0.5em} 
 \\
 \hspace{-1em}
  P(\nu_\mu \to \nu_\mu) &=&
     A_{\mu\mu} \cos\delta
  +  B_{\mu\mu} \sin\delta
  +  C_{\mu\mu}
  +  D_{\mu\mu} \cos2\delta
  +  E_{\mu\mu} \sin2\delta
, 
  \label{eq:general-result-prob-mu-mu-aa}
\\
A_{\mu\mu} &=&
  2  
   {\rm Re}[({S}'_{\mu \mu} c_{23}^2 + S'_{\tau \tau} s_{23}^2)^{*}
     ({S}'_{\tau \mu} + S'_{\mu \tau})] c_{23} s_{23} , \hspace{0.5em} 
 \\
B_{\mu\mu} &=&
 - 2  
  {\rm Im}[({S}'_{\mu \mu} c_{23}^2 + S'_{\tau \tau} s_{23}^2)^{*}
     ({S}'_{\tau \mu} - S'_{\mu \tau})] c_{23} s_{23}, \hspace{0.5em} 
 \\
C_{\mu\mu} &=&
    |S'_{\mu \mu}|^2   c_{23}^4
  + (|S'_{\mu \tau}|^2 + |S'_{\tau \mu}|^2) c_{23}^2 s_{23}^2
  + |S'_{\tau \tau}|^2   s_{23}^4
  + 2  {\rm Re}[{S}_{\mu \mu}^{'*} S'_{\tau \tau}] c_{23}^2 s_{23}^2 , \hspace{0.5em} 
\\
D_{\mu\mu} &=&
  2{\rm Re}[{S}_{\tau \mu}^{'*}S'_{\mu \tau}] c_{23}^2 s_{23}^2 , \hspace{0.5em} 
\\
E_{\mu\mu} &=&
  2{\rm Im}[{S}_{\tau \mu}^{'*}S'_{\mu \tau}] c_{23}^2 s_{23}^2 , \hspace{0.5em} 
  \label{eq:general-prob-mumu-aa}
\\
 \hspace{-1em}
  P(\nu_\tau \to \nu_\tau) &=&
     A_{\tau\tau} \cos\delta
  +  B_{\tau\tau} \sin\delta
  +  C_{\tau\tau}
  +  D_{\tau\tau} \cos2\delta
  +  E_{\tau\tau} \sin2\delta
, 
  \label{eq:general-result-prob-tau-tau-aa}
\\
A_{\tau\tau} &=& 
  - 2
   {\rm Re}[({S}'_{\mu \mu} s_{23}^2 + S'_{\tau \tau} c_{23}^2)^{*}
     ({S}'_{\tau \mu} + S'_{\mu \tau})] c_{23} s_{23} , \hspace{0.5em} 
 \\
B_{\tau\tau} &=& 
   2 
  {\rm Im}[({S}'_{\mu \mu} s_{23}^2 + S'_{\tau \tau} c_{23}^2)^{*}
     ({S}'_{\tau \mu} - S'_{\mu \tau})] c_{23} s_{23} , \hspace{0.5em} 
 \\
C_{\tau\tau} &=& 
    |S'_{\mu \mu}|^2 s_{23}^4
  + (|S'_{\mu \tau}|^2 + |S'_{\tau \mu}|^2) c_{23}^2 s_{23}^2
  + |S'_{\tau \tau}|^2   c_{23}^4
  + 2  {\rm Re}[{S}_{\mu \mu}^{'*} S'_{\tau \tau}] c_{23}^2 s_{23}^2 , \hspace{0.5em} 
 \\
D_{\tau\tau} &=& 
 2 {\rm Re}[{S}_{\tau \mu}^{'*}S'_{\mu \tau}] c_{23}^2 s_{23}^2 , \hspace{0.5em} 
\\
E_{\tau\tau} &=& 
 2 {\rm Im}[{S}_{\tau \mu}^{'*}S'_{\mu \tau}] c_{23}^2 s_{23}^2  , \hspace{0.5em} 
  \label{eq:general-prob-tautau-aa}
\\
 \hspace{-1em}
  P(\nu_\mu \to \nu_\tau) &=&
     A_{\mu\tau} \cos\delta
  +  B_{\mu\tau} \sin\delta
  +  C_{\mu\tau}
  +  D_{\mu\tau} \cos2\delta
  +  E_{\mu\tau} \sin2\delta
, 
  \label{eq:general-result-prob-mu-tau-aa}
\\
A_{\mu\tau} &=& 
  - 2  
  {\rm Re}[({S}'_{\mu \mu} - S'_{\tau \tau})^{*}
  ({S}'_{\tau \mu} c_{23}^2 - S'_{\mu \tau} s_{23}^2)]c_{23} s_{23} , \hspace{0.5em} 
 \\
B_{\mu\tau} &=& 
   2 
 {\rm Im}[({S}'_{\mu \mu} - S'_{\tau \tau})^{*}
 ({S}'_{\tau \mu} c_{23}^2 + S'_{\mu \tau} s_{23}^2)] c_{23} s_{23} , \hspace{0.5em} 
 \\
C_{\mu\tau} &=& 
    (|S'_{\mu \mu}|^2  + |S'_{\tau \tau}|^2)   c_{23}^2 s_{23}^2
  + |S'_{\mu \tau}|^2  s_{23}^4 + |S'_{\tau \mu}|^2 c_{23}^4 
  - 2 {\rm Re}[{S}_{\mu \mu}^{'*} S'_{\tau \tau}] c_{23}^2 s_{23}^2 , \hspace{0.5em} 
 \\
D_{\mu\tau} &=& 
  - 2 
  {\rm Re}[{S}_{\tau \mu}^{'*}S'_{\mu \tau}] c_{23}^2 s_{23}^2 , \hspace{0.5em} 
\\
E_{\mu\tau} &=& 
  - 2 
  {\rm Im}[{S}_{\tau \mu}^{'*}S'_{\mu \tau}] c_{23}^2 s_{23}^2, \hspace{0.5em} 
  \label{eq:general-prob-mutau-aa}
\end{eqnarray}
and the probabilities for the other channels
$P(\nu_\mu \to \nu_e)$, 
$P(\nu_\tau \to \nu_e)$ and 
$P(\nu_\tau \to \nu_\mu)$
are obtained by the replacements 
$S'_{\alpha\beta} \to S'_{\beta\alpha}$ and $\delta \to -\delta$
in $P(\nu_e \to \nu_\mu)$, 
$P(\nu_e \to \nu_\tau)$ and 
$P(\nu_\mu \to \nu_\tau)$, respectively. 
From these expressions,
we can see that matter effects 
is renormalized in $S'_{\alpha\beta}$, 
which does not contain CP phase $\delta$. 
Note that matter effects and 
CP effects are completely separated in the oscillation probability.
The mixing angle $\theta_{23}$
is also separated from matter effects and 
all of the oscillation probabilities are quartet polynomials of
$c_{23}$ and $s_{23}$.

\section{Coefficients in Symmetric Matter Profile}
\label{sec:appendix-b}
\hspace*{\parindent}
In this appendix, we give the relations of the coefficients 
of the probability in symmetric matter.
We use the condition 
\begin{eqnarray}
 \label{eq:sdashagain}
S'_{\alpha\beta}=S'_{\beta\alpha}, 
\end{eqnarray}
in symmetric matter 
profile.  
First, 
we calculate the relations between the coefficients
of T-conjugate probabilities. 
From the oscillation probabilities
(\ref{eq:general-result-prob-emu-aa}) and 
(\ref{eq:general-result-prob-mu-tau-aa})
in appendix \ref{sec:appendix-a}
and the symmetry of $S'$ (\ref{eq:sdashagain}), 
we obtain 
\begin{eqnarray}
  \label{eq:abc-s}
&&
A_{e \mu}-A_{\mu e} 
=
   4
  c_{23} s_{23}
 {\rm Re}[
 {S}_{\mu e}^{'*}
 S'_{\tau e}
-
 {S}_{e \mu}^{'*}
 S'_{e \tau}
]
=0, 
 \\
&&
B_{e \mu}
+ B_{\mu e}
=
  - 
  4
  c_{23} s_{23}
 {\rm Im}[
 {S}_{\mu e}^{'*}
 S'_{\tau e}
-
 {S}_{e \mu}^{'*}
 S'_{e \tau}
]
=0
, 
 \\
&&
C_{e \mu}
-C_{\mu e}
=
   (|S'_{\mu e}|^2 - |S'_{e \mu}|^2) c_{23}^2
 + (|S'_{\tau e}|^2| - |S'_{e \tau}|^2)  s_{23}^2
=0,
  \label{eq:abcd-s}
  \\
&&
A_{\mu\tau}-A_{\tau\mu} 
=
   4
 {\rm Re}[({S}'_{\mu \mu} - S'_{\tau \tau})^{*}
     (
 {S}'_{\tau \mu}
 -
 S'_{\mu \tau}
)
]
=0, 
 \\
&&
B_{\mu\tau}
+ B_{\tau\mu}
=
  - 
  4
 {\rm Im}[({S}'_{\mu \mu} - S'_{\tau \tau})^{*}
     (
 {S}'_{\tau \mu}
-
 S'_{\mu \tau})
]
=0
, 
  \label{eq:mutau-s-eq}
 \\
&&
C_{\mu\tau}-
C_{\tau\mu}
=
(
 |S'_{\mu \tau}|^2
-
 |S'_{\tau \mu}|^2
)
(
  s_{23}^4
 +
 c_{23}^4 
)
=0, 
 \\
&&
D_{\mu\tau}
-D_{\tau\mu}
=
-
 4
(|S'_{\mu \tau}|^2
-
|S'_{\tau \mu}|^2)
 c_{23}^2 s_{23}^2 
=0.
  \end{eqnarray}
The relations between the coefficients for 
$\nu_e \leftrightarrow \nu_\tau$ 
are obtained 
in the same way. 

Second, 
we calculate the coefficients of $\sin \delta$
and $\sin 2\delta$
in $P(\nu_\mu \to \nu_\mu)$ and $P(\nu_\tau \to \nu_\tau)$
and the coefficients of $\sin 2\delta$
in $P(\nu_\mu \to \nu_\mu)$ and $P(\nu_\tau \to \nu_\tau)$. 
From the concrete expression of oscillation probabilities
(\ref{eq:general-result-prob-mu-mu-aa}), 
(\ref{eq:general-result-prob-tau-tau-aa}) and
(\ref{eq:general-result-prob-mu-tau-aa})
in appendix \ref{sec:appendix-a}, 
we obtain 
\begin{eqnarray}
  \label{eq:be-ab}
&&
B_{\mu\mu}=
  - 
  2  c_{23} s_{23}
{\rm Im}[({S}'_{\mu \mu} c_{23}^2 + S'_{\tau \tau} s_{23}^2)^{*}
     ({S}'_{\tau \mu} - S'_{\mu \tau})
]
=0,
 \\
&&
E_{\mu\mu}=
  2
 {\rm Im}[{S}_{\tau \mu}^{'*}S'_{\mu \tau}] c_{23}^2 s_{23}^2 
=
  2
 {\rm Im}[|S'_{\mu \tau}|^2] c_{23}^2 s_{23}^2 
=0
,
 \\
&&
B_{\tau\tau}=
  2  c_{23} s_{23}
 {\rm Im}[({S}'_{\mu \mu} s_{23}^2 + S'_{\tau \tau} c_{23}^2)^{*}
     ({S}'_{\tau \mu} - S'_{\mu \tau})]
=0,
 \\
&&
E_{\tau\tau}=
  2  
 {\rm Im}[{S}_{\tau \mu}^{'*}S'_{\mu \tau}] c_{23}^2 s_{23}^2 
=
  2 
 {\rm Im}[|S'_{\mu \tau}|^2] c_{23}^2 s_{23}^2 
=0
,
 \\
&&
E_{\mu\tau} = 
- 2 
 {\rm Im}[{S}_{\tau \mu}^{'*}S'_{\mu \tau}] c_{23}^2 s_{23}^2 
=
- 2 
 {\rm Im}[|S'_{\mu \tau}|^2] c_{23}^2 s_{23}^2 
=0
,
 \\
&&
E_{\tau\mu} = 
 2 
 {\rm Im}[{S}_{\mu \tau}^{'*}S'_{\tau \mu}] c_{23}^2 s_{23}^2 
=
2 
 {\rm Im}[|S'_{\tau \mu}|^2] c_{23}^2 s_{23}^2 
=0
, 
  \end{eqnarray}
from the condition (\ref{eq:sdashagain}). 


\begin{thebibliography}{99}

\bibitem{solar} 
Homestake Collaboration, B.~T.~Cleveland {\rm et al.}, 
Astrophys. J. {\bf 496}, 505 (1998); 
SAGE Collaboration, J.~N.~Abdurashitov {\rm et al.}, 
Phys. Rev. {\bf C60} (1999) 055801; 
GALLEX Collaboration,  W.~Hampel {\rm et al.}, 
Phys. Lett. {\bf B447} (1999) 127; 
Super-Kamiokande Collaboration, Y. Fukuda  {\rm et al.}, 
Phys. Rev. Lett. {\bf 82} (1999) 1810; 
 {\rm ibid.}, 82 (1999) 2430;
SNO Collaboration, 
Phys. Rev. Lett. {\bf 89} (2002) 011301;
{\rm ibid.}, 
Phys. Rev. Lett. {\bf 89} (2002) 011302. 

\bibitem{atm} 
IMB Collaboration, 
R.~Becker-Szendy {\rm et al.}, 
 Phys. Rev. {\bf D46} (1992) 3720; 
SOUDAN2 Collaboration, 
W.W.M.~Allison 
{\rm et al.}, Phys. Lett. {\bf B391} (1997) 491; 
{\rm ibid.}, Phys. Lett. {\bf B449} (1999) 137; 
SuperKamiokande Collaboration, Y. Fukuda {\rm et al.}, 
 Phys. Rev. Lett. {\bf 82} (1999) 2644; 
{\rm ibid.}, 
Phys. Lett. {\bf B467} (1999) 185. 

\bibitem {CHOOZ}
CHOOZ Collaboration, M. Apollonio {\rm et al.}, , 
Phys. Lett. {\bf B466} (1999) 415. 

\bibitem{lbl}
   S.~Geer, Phys. Rev. {\bf D57} (1998) 6989
; Erratum-ibid. {\bf D59} (1999) 039903;
 OPERA Collaboration, K.~Kodama {\rm et al.}, 
 CERN/SPSC 99-20, SPSC/M635, LNGS-LOI 19/99 (1999); 
 ICARUS and NOE Collaborations, F.~Arneodo {\rm et al.}, 
 INFN/AE-99-17, CERN/SPSC 99-25, SPSC/P314 (1999); 
 MINOS Collaboration, P.~Adamson {\rm et al.}, 
 NuMI-L-337 (1998);
M. Aoki {\rm et al.}, 
hep-ph/0112338.


\bibitem{MSW}
   S.~P.~Mikheev and A.~Yu.~Smirnov, 
   Sov. J. Nucl. Phys. {\bf 42}, 913 (1985);
   L.~Wolfenstein, Phys. Rev. {\bf D17} (1978) 2369.  


\bibitem{Kuo}
T.K. Kuo and J. Pantaleone, Phys. Lett. {\bf B198}, (1987) 406. 

\bibitem{Minakata-Watanabe}
H. Minakata and S. Watanabe, 
Phys. Lett. {\bf B468}, (1999) 256.

\bibitem{Langacker}
P.Langacker, S.T. Petcov, G. Steigman and S. Toshev,
Nucl. Phys. {\bf B282} (1987) 589, 

\bibitem{AS}
Phys.Rev. D55 (1997) 1653-1658
J. Arafune and J. Sato, 
Phys. Rev. {\bf D55} (1997) 1653;
J. Arafune M. Koike and J. Sato, 
Phys. Rev. {\bf D56} (1997) 3093
; Erratum-ibid. {\bf D60} (1999)
     119905. 

\bibitem{Cervera}
A. Cervera, A. Donini, M.B. Gavela, J.J. Gomez Cadenas, 
P. Hernandez, O. Mena and S. Rigolin
Nucl. Phys. {\bf B579} (2000) 17; Erratum-ibid. {\bf B593} (2001) 731. 
 
\bibitem{Freund} M. Freund, Phys. Rev. {\bf D64},
(2001) 053003.

\bibitem{Krastev}
P.I. Krastev and S.T. Petcov, 
Phys. Lett. {\bf B205}, (1988) 84.


\bibitem{S0008056}
J. Sato, 
Nucl. Instrum. Meth. {\bf A472} (2000) 434.

\bibitem{Lipari} P. Lipari, Phys. Rev. {\bf D64}, (2001)

\bibitem{Yasuda} 
   O.~Yasuda, 
   Acta. Phys. Polon. {\bf B 30} (1999) 3089. 

\bibitem{Jarlskog}
 C. Jarlskog, Phys. Rev. Lett. {\bf 55} (1985) 1039. 

\bibitem{Naumov} 
V.A. Naumov, 
Int. J. Mod. Phys. {\bf D1} (1992) 379 

\bibitem{Harrison-Scott} 
   P.F.~Harrison and W.G.~Scott, 
   Phys. Lett. {\bf B476} (2000) 349. 

\bibitem{YKT} 
H. Yokomakura, K. Kimura and A. Takamura
, Phys. Lett. {\bf B496} (2000) 175.

\bibitem{Parke-Weiler} 
S.J. Parke and T.J. Weiler
Phys. Lett. {\bf B501} (2001) 106.

\bibitem{PREM}
 F. D. Stacey, Physics of the Earth, 2nd ed. (Wiley, 1977)

\bibitem{Minakata9705}
H. Minakata and H. Nunokawa, 
Phys. Rev. {\bf D57}, (1998) 4403.

\bibitem{KS98K2K}
M. Koike and J. Sato
Mod. Phys. Lett. {\bf A14} (1999) 1297.

\bibitem{OS}
T. Ota and J. Sato, 
Phys. Rev. {\bf D63} (2001) 093004.

\bibitem{Miura0106} 
T. Miura, T. Shindou, E. Takasugi and M. Yoshimura, 
Phys. Rev. {\bf D64} (2001) 073017.

\bibitem{Pinney} J. Pinney and O. Yasuda, Phys. Rev. {\bf D64},
 (2001) 093008.

\bibitem{Miura0102} 
T. Miura, E. Takasugi, Y. Kuno and M. Yoshimura, 
Phys. Rev. {\bf D64} (2001) 013002.

\bibitem{Akhmedov}
E.Kh. Akhmedov, P. Huber, M. Lindner and T. Ohlsson
Nucl. Phys. {\bf B608} (2001) 394. 

\bibitem{MNS}
Z.~Maki, M.~Nakagawa and S.~Sakata, 
 Prog. Theor. Phys. {\bf 28} (1962) 870. 

\bibitem{Zaglauer} H.W. Zaglauer and K.H. Schwarzer,
Z. Phys. {\bf C40}, (1988) 273. 

\bibitem{Kimura0203} K. Kimura, A. Takamura and H. Yokomakura,
Phys. Lett. {\bf B537} (2002) 86.

\bibitem{Kimura0205} K. Kimura, A. Takamura and H. Yokomakura,
hep-ph/0205295.

\bibitem{Minakata0204}
H. Minakata, H. Nunokawa and S. Parke, 
Phys.Lett. {\bf B537} (2002) 249

\bibitem{Minakata0108}
H. Minakata and H. Nunokawa, 
JHEP 0110 (2001) 001.

\bibitem{Fishbane}
P.M. Fishbane and P. Kaus, 
Phys. Lett. {\bf B495}, (2000) 369.

\end{thebibliography}
\end{document}